\documentstyle[prl,twocolumn,aps]{revtex}
\input{epsf}
\begin{document}
\draft
\twocolumn[\hsize\textwidth\columnwidth\hsize\csname @twocolumnfalse\endcsname
\title{Dynamical turbulent flow on the Galton board with friction}

\author{A. D. Chepelianskii$^{(a)}$ and D. L. Shepelyansky$^{(b)}$}

\address {$^{(a)}$ Lyc\'ee Pierre de Fermat, Parvis des Jacobins, 31068 
Toulouse Cedex 7, France}
\address {$^{(b)}$ Laboratoire de Physique Quantique, UMR 5626 du CNRS, 
Universit\'e Paul Sabatier, 31062 Toulouse Cedex 4, France}

\date{January 1, 2001}

\maketitle

\begin{abstract}
We study numerically and analytically the dynamics of particles on the 
Galton board, a regular lattice of disc scatters, in the presence of a 
constant external force and friction.
It is shown that under certain conditions friction leads to the
appearance of a strange chaotic attractor in an initially conservative 
Hamiltonian system. In this regime the particle flow becomes 
turbulent and its average velocity depends in  nontrivial manner on 
friction and other system parameters. We discuss the applications 
of these results to the transport properties of suspended particles in
a laminar viscous flow streaming through scatters.
\end{abstract}
\vskip1pc]

\narrowtext

It is well known that dissipation can lead to the appearance of strange
chaotic attractors in nonlinear nonautonomous dynamical systems 
\cite{lieberman,ott}. In this case the energy dissipation is compensated 
by an external energy flow so that stationary chaotic oscillations 
set in on the attractor. Such an energy flow is absent in the Hamiltonian
conservative systems and therefore the introduction of dissipation 
or friction is expected to drive the system to simple fixed points in 
the phase space. This rather general expectation is surly true 
if the system phase space is bounded. However a much richer situation 
appears in the case of unbounded space, where unexpectedly a strange 
attractor can be induced by dissipation in an originally conservative 
system. To investigate this situation we study the dynamics 
of particles on the Galton board in the presence of a constant external 
field and friction. This board, introduced by Galton in 1889 \cite{galton},
represents a triangular lattice of rigid discs with which particles 
collide elastically.  For the case of free particle motion, the collisions
with discs make the dynamics completely chaotic on the energy surface 
as it was shown by Sinai \cite{sinai}. In this paper we study 
how the dynamics is changed if the particles are affected by a 
friction force $\mathbf{F_f}$ which is assumed to be proportional to the 
particle velocity $\mathbf{v}$ and directed against it: 
$\mathbf{F_f} = - \gamma \mathbf{v}$.
In the free space without discs, an external field with an effective force 
$\mathbf{f}$ creates a stationary particle flow with the velocity 
$\mathbf{v_f} = \mathbf{f} / \gamma$. All perturbations decay to this 
flow with the rate proportional to $\gamma$ 
so that this laminar flow can be considered as a
simple attractor. It is interesting to understand how this flow will
be affected by the scattering on regularly placed discs.
As we will show latter on, in this case the flow can become turbulent and 
chaotic, and demonstrate many unusual properties (see Fig.1). 
The Hamiltonian dynamics of the model has been studied in great detail
(see \cite{gaspard} and Refs. therein) however, as for our knowledge,
the effects of friction were not addressed up to now.
The investigation of the dynamics in this regime 
is interesting from the viewpoint of general physical grounds. 
But we also note that physically such a 
situation can correspond to the motion of small particles in a viscous liquid 
flow which laminarly streams through large scatters that in our 
case have the form of discs. The same model can describe the electron 
dynamics in antidot superlattice which has been experimentally realized
in semiconductor heterostructures \cite{weiss}. In such structures the 
effects of classical chaos play an important role \cite{geisel} and the
effects of friction we discuss here can appear for relatively strong
electric fields.
\begin{figure}
\epsfxsize=8cm
\epsfysize=8cm
\epsffile{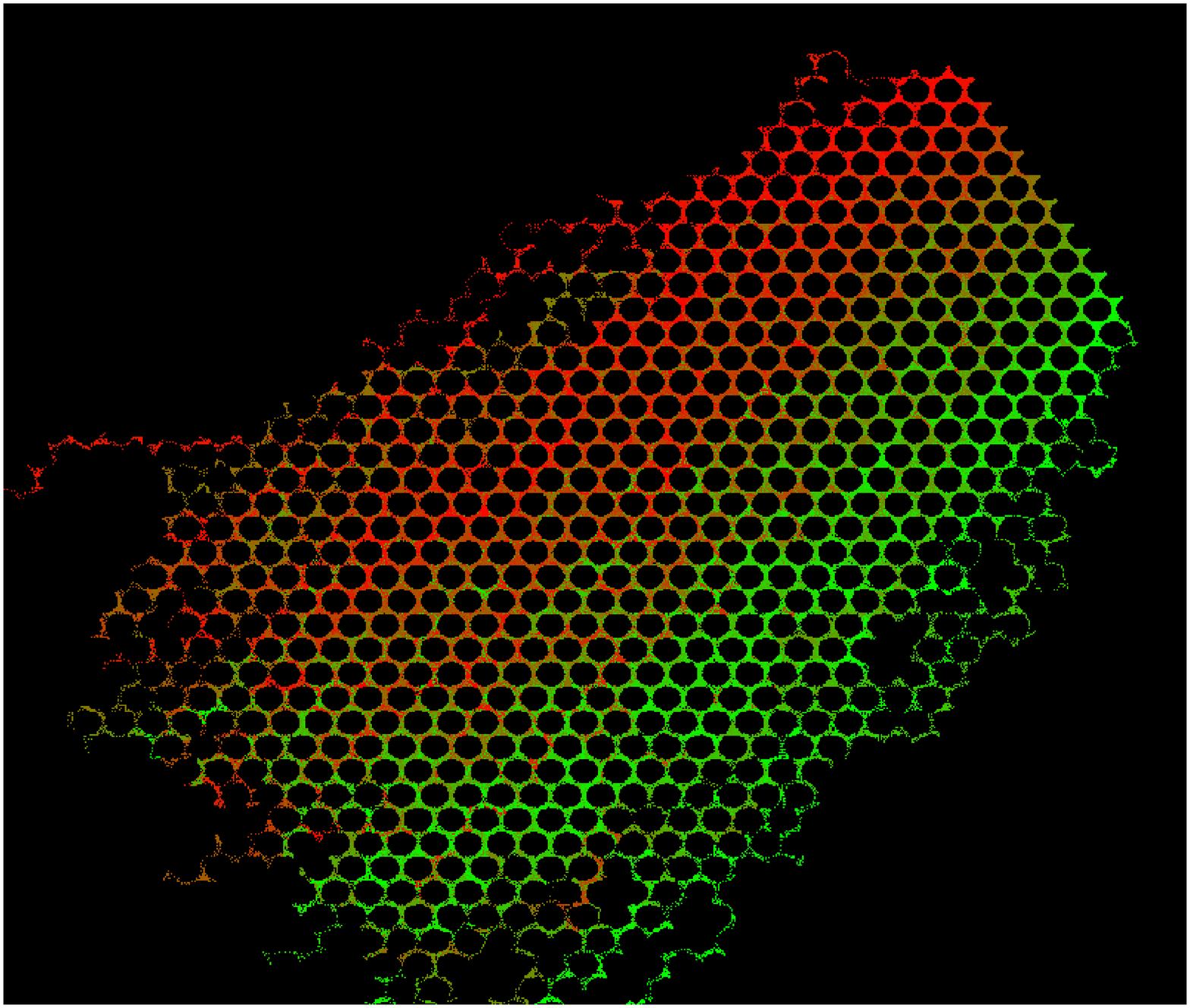}
\vglue 0.2cm
\caption{Chaotic flow on the Galton board. Here the distance 
between discs is $R = 2.24$, $\mathbf{f}$ = (-0.5, -0.5) and $\gamma = 0.1$.
Initially 100 particles are distributed homogeneously along 
a straight line segment in the upper right corner, their
color homogeneously changes from red to green along this segment.
} 
\label{fig1}
\end{figure}

To study the dynamics of this model, we fix the disc radius $a = 1$,
and the mass of the particles $m = 1$ so that the system is 
characterized only by three parameters: the distance $R$ between 
discs packed into equilateral triangles all over the $(x,y)$ plane;
the friction coefficient $\gamma$ and the external force of strength 
$f$. For $R \le R_c = 4/\sqrt{3}$ there is no straight line, by 
which a particle can cross the whole plane without collisions
(at $f = 0, \gamma = 0$), and we start the discussion from this case.
The particle dynamics is simulated numerically by using the exact 
solution of Newton equations between collisions, and by determining 
the collision points with the Newton algorithm. This way the 
trajectories are computed with high precision, and for example 
for $\gamma = 0$ the total energy is conserved with a relative 
precision of $10^{-14}$. A typical example of the chaotic flow
formed by an ensemble of particle trajectories is shown in Fig.~1.
To illustrate the mixing properties of the flow we attributed
a color to each trajectory that allows to follow their interpenetration
and spreading. This figure shows that there is a certain penetration
depth of one color into another, however this depth is finite since 
on average the initial color repartition is still visible.
\begin{figure}
\epsfxsize=7cm
\epsfysize=7cm
\epsffile{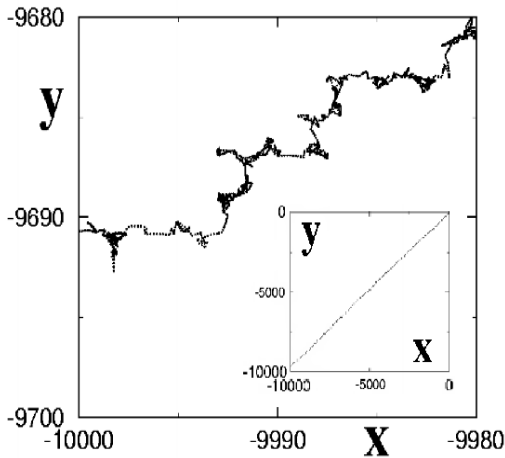}
\vglue 0.2cm
\caption{ Example of a single trajectory for the case of Fig.~1
shown on a small (main figure) and large (inset) scales.
The drift velocity of the flow is $v_f \approx 0.13$.
} 
\label{fig2}
\end{figure}

The properties of color penetration can be understood from the 
analysis of single trajectory dynamics. Such a typical example
is presented in Fig.~2. It shows that trajectory flows with
an average constant velocity $v_f$ in the direction of the 
external force. 
This drift velocity is constant only on average
since on a smaller scale the particle moves chaotically between
scatters (see Fig.~2). Even if the average flow is regular, 
on a small scale the particle dynamics is chaotic and since the 
dynamics is dissipative this means that the motion develops 
on a strange chaotic attractor. This attractor is induced 
by dissipation in a conservative Hamiltonian system.
For the case of Fig.~2 the drift velocity is relatively 
large and the particle does not have enough time to move around 
many scatters in the direction perpendicular to the flow. 
As a result the penetration depth for color mixing is not very large. 
However for smaller friction, the drift velocity becomes smaller 
and the penetration depth increases so that the particle makes many 
turns around discs as it is shown in Fig.~3.
Surprisingly, the drift velocity $v_f$ {\it increases } with the growth of the 
friction coefficient $\gamma $. This dependence is opposite to the case without
scatters where $v_f$ drops with the $\gamma$ growth.
\begin{figure}
\epsfxsize=7cm
\epsfysize=7cm
\epsffile{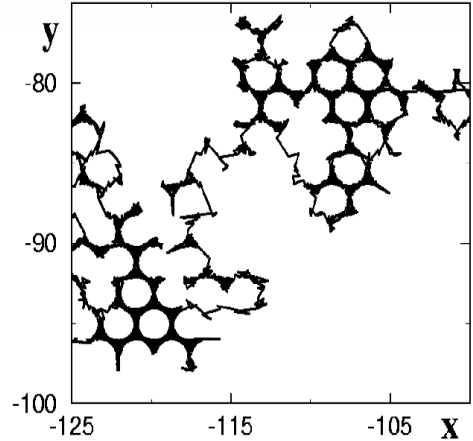}
\vglue 0.2cm
\caption{ Single trajectory for the same parameters as in Figs.~1,2 
but with smaller friction $\gamma = 0.004$; on a large scale the 
particle moves with a drift velocity $v_f \approx 0.05$ in 
the direction of external force.
} 
\label{fig3}
\end{figure}
\vglue -0.8cm
\begin{figure}
\epsfxsize=8cm
\epsfysize=6cm
\epsffile{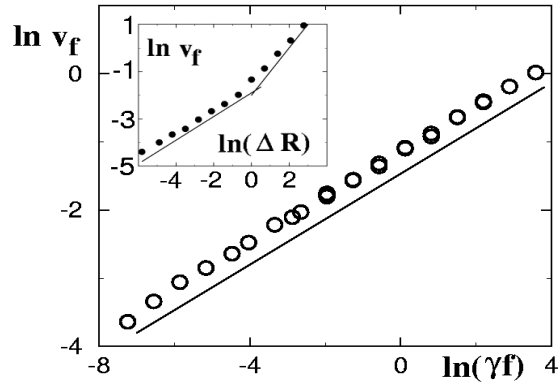}
\vglue 0.2cm
\caption{ Dependence of the flow velocity $v_f$ on the friction 
coefficient $\gamma$ and external field strength $f$ for $R = 2.24$,
$\mathbf{f} / |\mathbf{f}|$ = (-1, -1) and 
$0.001 \leq \gamma \leq 0.4; 0.5 \leq  f \leq 32$.
Numerical results are shown by circles, the straight line shows the 
dependence $v_f \propto (\gamma f)^{1/3}$.
Inset shows the dependence of $v_f$ on $\Delta R = R - 2$ for 
$\gamma = 0.1$ and $\mathbf{f}$ = (-0.5, -0.5): points give numerical 
data, lines show the slopes $0.5$ and $1$.
} 
\label{fig4}
\end{figure}

The dependence of the drift velocity $v_f$ on $\gamma$ and on the 
external force $f = |\mathbf{f}|$ is obtained by averaging over many
trajectories. Each trajectory is followed for a sufficiently long time 
so that the distance from the origin becomes larger than a few thousands
units. Up to fluctuations, the flow is always directed along 
the applied force $\mathbf{f}$.
The resulting dependence is shown in Fig.~4 according to which 
$v_f \propto (\gamma f)^{1/3}$. To understand the origin of this 
unexpected dependence on parameters we also compute the average 
velocity square $v^{2} = <v^2_x + v^2_y>$ and determine how it varies 
with $f$ and $\gamma$. The data presented in Fig.~5 give a dependence 
of the form: $v^2 \propto f^{4/3}/{\gamma}^{2/3}$. We stress that 
$v_f$ and $v^2$ are independent of initial velocity of the particle.

To explain the observed scaling relations for the dynamical turbulent
flow on the Galton board we give the following theoretical arguments.
In the regime with weak friction the particles start to diffuse 
among discs in a chaotic manner with the diffusion rate 
$D = v l/2$ where $v$ is the velocity of the particle and $l$ 
is the mean free path \cite{ziman}.
For $R \sim 1$ we have  $l \sim R \sim 1$
while the dependence of $l$ on $R$ will be discussed in more detail
latter. During the dissipative time scale $\tau_{\gamma} = m / \gamma$
this diffusion leads to the particle displacement 
$\Delta r \sim \sqrt{D m/ \gamma}$
along the direction of applied field. Here, for generality we introduced
the particle mass $m$. This gives the change in the 
potential energy $U \sim f \Delta r \sim f \sqrt{D m/ \gamma}$.
In the stationary regime at time $t \gg \tau_{\gamma}$, this potential
energy should be comparable with the kinetic energy of the particle so
that $U \sim m v^2$. As a result we find that the average velocity square 
is 
\begin{equation}
\label{velav}
v^2 \sim f^{4/3} (l / \gamma m)^{2/3} \;\; ,
\end{equation}
that is in agreement with the data of Fig.~5.
\begin{figure}
\epsfxsize=8cm
\epsfysize=6cm
\epsffile{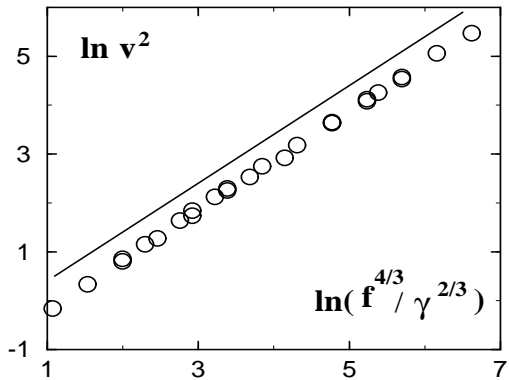}
\vglue 0.2cm
\caption{ Dependence of average $v^2$ on $f$ and $\gamma$ for 
$R = 2.24$, 
$\mathbf{f} / |\mathbf{f}|$ = (-1, -1) and parameter range 
$0.001 \leq \gamma \leq 0.4; 0.5 \leq  f \leq 32$.
Circles show numerical data, the straight line gives the slope $1$.
} 
\label{fig5}
\end{figure}

The relation (\ref{velav}) allows to determine the mobility $\mu$ which 
relates the velocity of the flow with the applied force: 
$\mathbf{v_f} = \mu \; \mathbf{f}$. Indeed during the time $\tau_c$ 
between collisions the particle is accelerated by force that gives 
the average drift velocity $v_f = f \tau_c / m$ and $\mu = \tau_c / m$.
Since the dynamics is chaotic the direction of 
velocity is changed randomly after each collision 
so that $v_f$ is accumulated only between collisions. 
The time $\tau_c$ is determined by the mean 
free path $l$ and the average velocity $v \; :$
$\tau_c \; = l / v \; = 2 D / v^2$ and $\mu = \; D / (m v^2/2)$. 
In fact this is the Einstein relation 
according to which the mobility is given by ratio of the diffusion 
rate to the average kinetic energy (temperature) \cite{landau}. 
Together with (\ref{velav}) this gives the drift velocity of 
the chaotic flow:
\begin{equation}
\label{veldrift}
v_f \sim l^{2/3} (f \gamma /m^2)^{1/3} \;\;,
\end{equation}
which is in a good agreement with the numerical results in Fig.~4.
On average the particle moves with the velocity $v_f$ in the direction
of the force (see an example in the inset of Fig.~2). The amplitude
of fluctuations around this direction is $\Delta r \sim \sqrt{D m / \gamma}$ 
which also determines the color mixing depth (see Fig.~1).

The relations (\ref{velav}), (\ref{veldrift}) allow to estimate 
the value of the Lyapunov exponent $\lambda$. Indeed, the particle 
moves with a typical velocity $v$ and as in the case of the Sinai 
billiard with $l \sim R \sim 1$ we have $\lambda \sim v / l$.
Therefore  in the regime when:
\begin{equation}
\label{gammac}
\gamma < \gamma_c \sim \sqrt{f m / l} \;\; ,
\end{equation}
the value of $\lambda$ is much larger than the dissipation rate $\gamma / m$.
As a result for $\gamma \ll \gamma_c$ the strange attractor is fat 
and its fractal dimension is close to the maximal dimension $4$,
which is determined by the number of degrees of freedom (we remind 
that contrary to the nondissipative case the energy is not conserved).

The physical meaning of the critical $\gamma_c$ border is the following:
for $\gamma \gg \gamma_c$ the dissipation time $\tau_{\gamma}$ becomes 
much shorter than the time between collisions $\tau_c$. In this case 
the dissipation dominates chaos and the strange attractor degenerates into
a simple attractor. Among such simple attractors we observed trajectories 
that move regularly under some angle to the applied force but more often
trajectories that simply drop to a fixed point on a disc where the applied 
force is perpendicular to the disc border. Of course the numerical 
simulations can not guarantee that the chaotic attractor at 
$\gamma < \gamma_c$ will not degenerate to a simple one at very long 
times. However all our simulations performed with high computer 
precision give no indications in this direction even when we followed 
trajectories up to very far distances from the origin (bigger than $10^5 $
length units). 

The equation (\ref{gammac}),
which determines the border for disappearance  of chaotic flow,
is written for the case $l \sim R \sim 1$.
Therefore, it should be modified when $R \gg 1$ and $\Delta R = R - 2 \ll 1$.
Indeed for $R \gg 1$ the mean free path $l$ is given by 
$l = 1 / (\sigma n)$ where $\sigma = 2 a = 2$ is the disc cross section 
and $n = 2 / (\sqrt{3} R^2) $ is the density of scatters.
Then the time between collisions is $l / v$ and from equation (\ref{velav})
we obtain $\gamma_c \sim m v/l \sim \sqrt{ f m a } / R$. In the other limit
$\Delta R \ll 1$ we may assume that $D \sim v \Delta R$, $l \sim \Delta R$ 
and the escape time from a cell between three discs is 
$\tau_{esc} \sim a^2 / D$.
Therefore the dissipation becomes dominant and the chaotic  attractor
disappears at 
$\gamma > \gamma_c \sim m / \tau_{esc} \sim \sqrt{m f/ a^3} \; \Delta R$.

The above changes in the mean free path $l$ also affect the drift 
velocity of the flow through the relation (\ref{veldrift}).
Indeed for $\Delta R \ll 1$ we expect $l \sim \Delta R$ that 
gives $v_f \propto {\Delta R}^{2/3}$. This dependence is close 
to the numerical data shown in Fig.~4 (inset) even if the numerical 
value of the exponent is approximated better by $0.5$.
In the other limit $R \gg 1$, we have $l \sim R^2 / a$ that gives 
$v_f \propto R^{4/3}$. This power dependence is in a satisfactory 
agreement with our numerical data (inset in Fig.~4) although 
the numerical value of the exponent is closer to $1$.
We attribute these small deviations in the exponent values to 
a restricted interval of variation in $R$. Actually we can not 
use very large/small values of $\Delta R$ since in these limits 
the value of $\gamma$ becomes comparable with $\gamma_c$
and chaotic attractor disappears. We note that according to our 
data (see Fig.~4) the strange attractor exists even in the case 
$R > R_c = 4/\sqrt{3}$ when at $f = 0, \gamma = 0$ there are straight
trajectories crossing the whole plane without collision.
Apparently the contribution of these orbits is not significant 
if $\gamma > 0$ and $\mathbf{f}$ is not directed along these lines.

Let us now discuss a few consequences of the obtained results. 
The introduction of scatters qualitatively changes the flow 
properties: instead of a laminar flow we obtain a turbulent/chaotic
one. The flow rate $v_f$ depends on the applied force $f$ and $\gamma$
in a nonlinear way given by equation (\ref{veldrift}). Contrary
to the laminar case where $v_f = f / \gamma$ in presence of scatters
the flow velocity depends also on the mass of particles $m$.
In fact light particles stream with higher velocity. This property 
of chaotic flow can be used for mass separation of particles suspended in 
viscous flows, for example isotopes. It is also interesting to note that 
in the derivation of equations 
(\ref{velav}),~(\ref{veldrift}),~(\ref{gammac})
we didn't use any specific properties of the disc distribution in the 
plane. Therefore these results remain also valid for a random distribution
of disc scatters in the plane. We wonder if some rigorous mathematical 
results can be obtained in the limit of weak friction
to prove the existence of a chaotic attractor
for a regular or random distribution of discs.

We think that the obtained results represent certain interest for
transport properties of particles suspended in a viscous flow 
streaming through a system of large scatters. Indeed, when a 
liquid streams laminarly through scatters with the velocity $v_{s}$
the flow of suspended particles is characterized by the
relations (\ref{velav}),~(\ref{veldrift}),~(\ref{gammac}) with 
$f_{eff} = v_{s} {\gamma}_{eff}$ where ${\gamma}_{eff}$ is an effective
friction created by the viscosity of the liquid and $f_{eff}$ is the 
resulting effective force acting on the suspended particles.
Then for weak friction the drift velocity of particles $v_f$ is smaller 
than the velocity of the liquid $v_s$.
Such kind of transport can be studied experimentally with viscous
liquids and its investigation can contribute to a better understanding 
of the interplay between dissipation, turbulence and chaos.

\end{document}